\documentclass[11pt,a4paper]{article}

\usepackage{amssymb}
\usepackage[dvips]{graphicx}
\usepackage{bm}

\begin{document}

\title{A class of the NSVZ renormalization schemes for ${\cal N}=1$ SQED}

\author{
I.O.Goriachuk\\
{\small{\em Moscow State University,}}\\
{\small{\em Faculty of Physics, Department of Theoretical Physics,}}\\
{\small{\em 119991, Moscow, Russia}},\\
\\
A.L.Kataev\\
{\small{\em Institute for Nuclear Research of the Russian Academy of Science,}}\\
{\small{\em 117312, Moscow, Russia}};\\
{\small{\em Moscow Institute of Physics and Technology,}}\\
{\small{\em 141700, Dolgoprudny, Moscow Region, Russia}},\\
\\
K.V.Stepanyantz\\
{\small{\em Moscow State University,}}\\
{\small{\em Faculty of Physics, Department of Theoretical Physics,}}\\
{\small{\em 119991, Moscow, Russia}}}

\maketitle

\begin{abstract}
For the ${\cal N}=1$ supersymmetric electrodynamics we investigate renormalization schemes in which the NSVZ equation relating the $\beta$-function to the anomalous dimension of the matter superfields is valid in all loops. We demonstrate that there is an infinite set of such schemes. They are related by finite renormalizations which form a group and are parameterized by one finite function and one arbitrary constant. This implies that the NSVZ $\beta$-function remains unbroken if the finite renormalization of the coupling constant is related to the finite renormalization of the matter superfields by a special equation derived in this paper. The arbitrary constant corresponds to the arbitrariness of choosing the renormalization point. The results are illustrated by explicit calculations in the three-loop approximation.
\end{abstract}

\unitlength=1cm

\vspace*{-17.4cm}

\begin{flushright}
INR-TH-2018-017
\end{flushright}

\vspace*{16.0cm}

\section{Introduction}
\hspace*{\parindent}

Supersymmetric quantum field theory models have many interesting properties. For example, supersymmetry leads to some non-renormalization theorems, which sometimes produce very non-trivial relations between various renormalization constants. In the case of ${\cal N}=1$ supersymmetry as one of the non-renormalization theorems it is possible to consider the so-called exact Novikov, Shifman, Vainshtein, and Zakharov (NSVZ) $\beta$-function \cite{Novikov:1983uc,Jones:1983ip,Novikov:1985rd,Shifman:1986zi}. For the pure ${\cal N}=1$ supersymmetric Yang--Mills (SYM) theory it gives the all-order expression for the $\beta$-function, while for ${\cal N}=1$ supersymmetric gauge theories with matter it relates the $\beta$-function to the anomalous dimensions of the matter superfields.  The exact NSVZ $\beta$-function was first derived for ${\cal N}=1$ non-Abelian gauge theories as a result of analysing the structure of instanton contributions \cite{Novikov:1983uc,Novikov:1985rd} (for a review see \cite{Shifman:1999mv}) and anomalies \cite{Jones:1983ip,Shifman:1986zi} (see also Ref. \cite{ArkaniHamed:1997mj}). In this paper we will consider the ${\cal N}=1$ supesymmetric quantum electrodynamics (SQED) with $N_f$ flavors, for which the NSVZ equation is written as \cite{Vainshtein:1986ja,Shifman:1985fi}

\begin{equation}\label{eq:NSVZ_equation}
\widetilde\beta(\alpha) = \frac{{\alpha}^2 N_f}{\pi}\Big(1 - \widetilde\gamma(\alpha)\Big).
\end{equation}

\noindent
Note that writing the NSVZ equation it is necessary to distinguish between the renormalization group functions (RGFs) defined in terms of the renormalized coupling constant ($\widetilde\beta(\alpha)$ and $\widetilde\gamma(\alpha)$) and the ones defined in terms of the bare coupling constant ($\beta(\alpha_0)$ and $\gamma(\alpha_0)$) \cite{Kataev:2013eta}. In particular, Eq. (\ref{eq:NSVZ_equation}) is written for the former RGFs, because below in this paper we will mostly deal with them.

It is known \cite{Stepanyantz:2011jy, Stepanyantz:2014ima} that for Abelian supersymmetric theories RGFs defined in terms of the bare coupling constant $\alpha_0$ satisfy the NSVZ relation in all orders in the case of using the higher derivative (HD) regularization \cite{Slavnov:1971aw,Slavnov:1972sq} in the supersymmetric version \cite{Krivoshchekov:1978xg,West:1985jx}. (The three-loop calculation confirming this fact can be found, e.g., in \cite{Kazantsev:2014yna}.) However, RGFs defined in terms of the renormalized coupling constant depend on the renormalization prescription. Namely, for single-charge theories the $\beta$-function becomes scheme-dependent starting from the three-loop approximation, while the anomalous dimension of the matter superfields is scheme-dependent starting from the two-loop approximation. That is why they in general do not satisfy the relation (\ref{eq:NSVZ_equation}) and it is necessary to use a special renormalization prescription(s) to obtain it. The subtraction schemes for which the NSVZ relation is valid are usually called ``the NSVZ schemes''.

In the case of using the higher derivative regularization the NSVZ relation in the Abelian case is satisfied in all orders if the renormalization prescription is constructed by imposing the boundary conditions \cite{Kataev:2013eta,Kataev:2013csa}

\begin{equation}\label{eqs:bc_HD_NSVZ}
\alpha_{0}(\alpha, x_{0}) = \alpha, \qquad Z(\alpha, x_{0}) = 1,
\end{equation}

\noindent
where $x_0$ is a fixed finite value of $\ln\Lambda/\mu$. This fact has been confirmed by explicit three-loop calculations, see, e.g., \cite{Kataev:2014gxa}. It is convenient to choose $x_0$ equal to 0, because in this case only powers of $\ln\Lambda/\mu$ are included into the renormalization constants. That is why this scheme can be called $\mbox{HD}+\mbox{MSL}$, where MSL is the abbreviation of ``Minimal Subtraction of Logarithms'' \cite{Shakhmanov:2017soc,Stepanyantz:2017sqg}. It has been proved that the $\mbox{HD}+\mbox{MSL}$ prescription gives the NSVZ-like schemes in all loops for the Adler $D$-function in ${\cal N}=1$ SQCD \cite{Shifman:2014cya,Shifman:2015doa,Kataev:2017qvk} and for the renormalization of the photino mass in softly broken ${\cal N}=1$ SQED \cite{Nartsev:2016nym,Nartsev:2016mvn}. Also there are strong evidences that the $\mbox{HD}+\mbox{MSL}$ prescription gives the NSVZ scheme in all orders for non-Abelian supersymmetric gauge theories \cite{Stepanyantz:2016gtk} regularized by higher covariant derivatives. Some recent explicit three-loop calculations \cite{Shakhmanov:2017soc,Kazantsev:2018nbl} confirm this guess.

However, supersymmetric theories are mostly regularized with the help of dimensional reduction \cite{Siegel:1979wq}. Effectively this method is reduced to dealing with $\gamma$-matrices and supersymmetric covariant derivatives in 4 dimensions and calculating loop integrals in $d$ dimensions. Most calculations with dimensional reduction were made in the $\overline{\mbox{DR}}$ scheme which is analogous to the $\overline{\mbox{MS}}$ scheme for the dimensional regularization. In this case the NSVZ equation is not valid already in the lowest order where the scheme dependence is essential (namely, for the three-loop $\beta$-function and the two-loop anomalous dimension) \cite{Jack:1996vg}. Nevertheless, it is possible to make a finite redefinition of the coupling constant which restores the NSVZ relation. This finite renormalization should be tuned in each order of the perturbation theory. In the three- and four-loop approximations it has been constructed in Refs. \cite{Jack:1996vg} and \cite{Jack:1996cn,Jack:1998uj}, respectively (see also Ref. \cite{Mihaila:2013wma} for a review.) However, at present there is no general all-order prescription for constructing this coupling constant redefinition relating the $\overline{\mbox{DR}}$ scheme to the NSVZ scheme. It is worth to note that in the three-loop approximation the NSVZ scheme constructed in this way for ${\cal N}=1$ SQED has also been formulated  \cite{Aleshin:2016rrr} using the boundary conditions similar to (\ref{eqs:bc_HD_NSVZ}).

Note that RGFs in the NSVZ scheme obtained with the HD regularization and RGFs in the NSVZ scheme obtained with dimensional reduction after redefinition of the coupling constant are different. This implies that these two NSVZ schemes are different, although the NSVZ relation is valid in both of them. Therefore, the validity of the NSVZ relation is not sufficient for fixing the subtraction scheme unambiguously. In this paper we demonstrate that there is a class of the NSVZ schemes parameterized by one function and one constant (which reflects the arbitrariness of choosing the normalization point $\mu$).

The paper is organized as follows: In Sect. \ref{sec:NSVZclass} we investigate a finite renormalization relating two arbitrary NSVZ schemes. It is shown that such a finite renormalization should satisfy a certain equation, which is exact in all orders and specifies a class of the NSVZ schemes. In Sect. \ref{sec:mu_change} we demonstrate that changing the renormalization point $\mu$ corresponds to a finite renormalization which does not break the NSVZ relation. In the three-loop approximation the general form of the finite renormalization preserving the NSVZ equation is constructed in Sect. \ref{sec:three_loop_NSVZclass}. A particular case of this finite renormalization corresponding to changing the renormalization point in the three-loop approximation is analysed in Sect. \ref{sec:three_loop_mu_change}. In Sect. \ref{sec:examples} we illustrate the above results by relating two different NSVZ schemes. One of them is obtained starting from the $\overline{\mbox{DR}}$ result by a finite renormalization (only) of the coupling constant, and the other is obtained with the help of the higher derivative regularization supplemented by the boundary conditions (\ref{eqs:bc_HD_NSVZ}).

\section{Finite renormalizations which do not break the NSVZ relation}
\hspace*{\parindent}\label{sec:NSVZclass}

Usually RGFs are defined in terms of the renormalized coupling constant,

\begin{equation}\label{eqs:RG}
\widetilde\beta(\alpha) \equiv \left.\frac{d\alpha\left(\alpha_0, \Lambda/\mu\right)}{d\ln\mu}\right\arrowvert_{\alpha_0=\mbox{\scriptsize const}};\qquad
\widetilde\gamma(\alpha)
\equiv \left.\frac{d\ln{Z}\left(\alpha\left(\alpha_0,\Lambda/\mu\right), \Lambda/\mu\right)}{d\ln\mu}\right\arrowvert_{\alpha_0=\mbox{\scriptsize const}},
\end{equation}

\noindent
where the bare coupling constant $\alpha_0$ and the renormalization point $\mu$ are considered as independent variables. In particular, this implies that in the equation for $\widetilde\gamma$ the derivative also acts on
$\ln\mu$ inside the function $\alpha(\alpha_0,\Lambda/\mu)$. Thus defined RGFs are scheme-dependent. Really, it is possible to make the finite renormalization of the coupling constant and of the matter superfields of the form

\begin{equation}\label{eq:fin_ren}
\alpha^\prime\left(\alpha_0,~\Lambda/\mu\right) = \alpha^\prime\left(\alpha\left(\alpha_0,~
\Lambda/\mu\right)\right);\qquad
Z^\prime\left(\alpha^\prime\left(\alpha\right),\Lambda/\mu\right) = z\left(\alpha\right)
Z\left(\alpha,\Lambda/\mu\right),
\end{equation}

\noindent
where $\alpha^\prime(\alpha)$ and $z(\alpha)$ are arbitrary finite functions. Under this finite renormalization RGFs (\ref{eqs:RG}) change as \cite{Vladimirov:1979my}

\begin{eqnarray}\label{eq:fin_ren_beta}
&& \widetilde\beta^\prime\left(\alpha^\prime(\alpha)\right) =
\frac{d\alpha^\prime(\alpha)}{d\alpha}\, \widetilde\beta(\alpha), \\
\label{eq:fin_ren_gamma}
&& \widetilde\gamma^\prime\left(\alpha^\prime(\alpha)\right) =
\frac{d\ln z(\alpha)}{d\alpha}\, \widetilde\beta(\alpha) +
\widetilde\gamma(\alpha).
\end{eqnarray}

\noindent
Let us consider a renormalization scheme for which the NSVZ relation ($\ref{eq:NSVZ_equation}$) is satisfied. Then we would like to find under what finite renormalizations ($\ref{eq:fin_ren}$) the NSVZ relation remains valid.
Assuming that Eq. (\ref{eq:NSVZ_equation}) is satisfied for the new RGFs ($\ref{eq:fin_ren_beta}$) and ($\ref{eq:fin_ren_gamma}$) it is easy to obtain \cite{Kutasov:2004xu}

\begin{equation}\label{eq:newNSVZ}
\frac{d\alpha^\prime}{d\alpha} \widetilde\beta(\alpha) =
\frac{{\alpha^\prime}^2 N_f}{\pi}\Big(1 - \frac{d\ln z(\alpha)}{d\alpha}\,
\widetilde\beta(\alpha) - \widetilde\gamma(\alpha)\Big).
\end{equation}

\noindent
The anomalous dimension can be found from the NSVZ relation,

\begin{equation}\label{eq:gmma_from_NSVZ}
\widetilde\gamma(\alpha) = 1 - \frac{\pi}{N_f} ~ \frac{\widetilde\beta(\alpha)}{\alpha^2}.
\end{equation}

\noindent
Substituting this expression into Eq. (\ref{eq:newNSVZ}) we obtain the differential equation for the finite functions $\alpha^\prime(\alpha)$ and $z(\alpha)$

\begin{equation}
\frac{d\alpha^\prime}{d\alpha} - \frac{{\alpha^\prime}^2}{\alpha^2} +
\frac{{\alpha^\prime}^2 N_f}{\pi} ~ \frac{d\ln z\left(\alpha\right)}{d\alpha} = 0
\end{equation}

\noindent
with the solution

\begin{equation}\label{eq:NSVZ_cons_fin_ren}
\frac{1}{\alpha^\prime(\alpha)} - \frac{1}{\alpha}
- \frac{N_f}{\pi}\ln{z(\alpha)} = B,
\end{equation}

\noindent
where $B = \mbox{const}$. Any finite renormalization of the form (\ref{eq:fin_ren}) which does not break the NSVZ relation should satisfy this equation. Therefore, the finite functions $\alpha^\prime(\alpha)$ and $z(\alpha)$ are no longer independent. We can consider, e.g., the function $z(\alpha)$ and the constant $B$ as independent parameters, but then the function $\alpha^\prime(\alpha)$ will be expressed in terms of them. Note that the finite constant $B$ appears in Eq. (\ref{eq:NSVZ_cons_fin_ren}) due to the arbitrariness of choosing a renormalization scale $\mu$ \cite{Vladimirov:1979my}. We will prove this explicitly in the next section.

The existence of the finite renormalizations which do not break the NSVZ equation implies that the NSVZ scheme is not unique. There is a class of such renormalization schemes parameterized by one finite function and one finite constant (e.g., $z(\alpha)$ and $B$).

Note that the sequence of two finite renormalizations $\{\alpha_1(\alpha),z_1(\alpha)\}$ and $\{\alpha_2(\alpha_1),z_2(\alpha_1)\}$ satisfying Eq. (\ref{eq:NSVZ_cons_fin_ren}) with the parameters $B^{(1)}$ and $B^{(2)}$, respectively, also satisfies this equation,

\begin{equation}
\frac{1}{\alpha_2(\alpha_1(\alpha))} - \frac{1}{\alpha} - \frac{N_f}{\pi}\ln\Big(z_2(\alpha_1(\alpha))z_1(\alpha)\Big) = B^{(1)} + B^{(2)}.
\end{equation}

\noindent
Therefore, the overall transformation can be written as

\begin{equation}\label{eq:two_transformations}
\alpha'(\alpha) = \alpha_2(\alpha_1(\alpha));\qquad z(\alpha) = z_2(\alpha_1(\alpha)) z_1(\alpha);\qquad B = B^{(1)} + B^{(2)}.
\end{equation}

\noindent
This implies that the finite renormalizations which do not change the NSVZ relation form a group.\footnote{Transitivity and reflexivity are usual properties of finite renormalizations, see, e.g., Refs. \cite{Brodsky:2012ms,Scimemi:2018xaf}.}

\section{Changing the renormalization scale $\mu$ within the class of NSVZ schemes}
\hspace*{\parindent}\label{sec:mu_change}

The dependence of the coupling constant $\alpha$ and of the renormalization constant $Z$ on the renormalization scale $\mu$ are described by RGFs (\ref{eqs:RG}) (see Ref. \cite{Vladimirov:1979ib} for details). Evidently, after the change of the normalization point from $\mu$ to $\mu^\prime = C\mu$, where $C = \mbox{const}$, RGFs (\ref{eqs:RG}) remain the same. Therefore, the NSVZ relation also remains unchanged under this transformation. However, as we will demonstrate in this section, such a rescaling changes a value of the constant $B$ in Eq. (\ref{eq:NSVZ_cons_fin_ren}). In particular, it can be completely removed by a special modification of $\mu$.

Let us find a finite renormalization corresponding to the change of the renormalization scale $\mu \to \mu^\prime$ assuming that $\alpha \to \alpha^\prime\left(\alpha\right)$ and $Z\left(\alpha, \Lambda/\mu\right) \to Z^\prime\left(\alpha^\prime, \Lambda/\mu^\prime\right) = z\left(\alpha\right)Z\left(\alpha,\Lambda/\mu\right)$. Integrating the renormalization group equations from $\mu$ to $\mu'$ we obtain

\begin{eqnarray}\label{eq:mu_chng_alpha_sol}
&& \ln{\frac{\mu^\prime}{\mu}} = \int\limits_{\alpha}^{\alpha^\prime(\alpha)}
\frac{d\alpha}{\widetilde\beta(\alpha)}, \\
\label{eq:mu_chng_z_sol}
&& z(\alpha) = \exp\Bigg(\int\limits_{\alpha}^{\alpha^\prime(\alpha)}
\frac{\widetilde\gamma(\alpha)}{\widetilde\beta(\alpha)} d\alpha\Bigg).
\end{eqnarray}

Let us solve the NSVZ relation (\ref{eq:NSVZ_equation}) for the anomalous dimension of the matter superfields $\widetilde\gamma$ and substitute the result (given by Eq. (\ref{eq:gmma_from_NSVZ})) into Eq. (\ref{eq:mu_chng_z_sol}). Then it is possible to take the integral with the help of Eq. (\ref{eq:mu_chng_alpha_sol}),

\begin{equation}
z(\alpha) = \exp{\Bigg(\ln{\frac{\mu^\prime}{\mu}} + \left.\frac{\pi}{N_f}
\frac{1}{\alpha}\right|_{\alpha}^{\alpha^\prime\left(\alpha\right)}\Bigg)}.
\end{equation}

\noindent
Substituting this expression into Eq. (\ref{eq:NSVZ_cons_fin_ren}) we obtain that the parameters of a finite renormalization corresponding to a change of the renormalization point $\mu\to\mu^\prime$ for the NSVZ schemes satisfy the equation

\begin{equation}
\frac{1}{\alpha^\prime(\alpha)} - \frac{1}{\alpha} -
\frac{N_f}{\pi} \ln z(\alpha) = -\frac{N_f}{\pi} \ln \frac{\mu^\prime}{\mu} = B =
\textrm{const}, \label{eq:NSVZ_B_hide}
\end{equation}

\noindent
which is exact in all orders. Thus, we see that the constant $B$ really corresponds to an arbitrariness in changing the renormalization point $\mu$ and (according to Eq. (\ref{eq:two_transformations})) can be removed by the transformation

\begin{equation}\label{eq:new-mu}
\mu \to \mu' = \mu \exp\Big(\frac{\pi}{N_f} B\Big).
\end{equation}

Therefore, the class of NSVZ schemes can be parameterized by a single finite function if the renormalization schemes which differ only by a value of $\mu$ are considered as equivalent. For example, as a parameter it is possible to choose the function $z(\alpha)$.

\section{Verification in the lowest orders}
\hspace*{\parindent}\label{sec:three_loop_NSVZclass}

The exact Eq. (\ref{eq:NSVZ_cons_fin_ren}) describes the class of finite renormalizations which do not break the NSVZ relation. In this section we explicitly verify that it is really so in the lowest orders of the perturbation theory. For this purpose we present RGFs as the series in the coupling constant,

\begin{eqnarray}\label{eqs:loop_exp_RG}
&& \widetilde\beta(\alpha) = \frac{\alpha^2}{\pi} \beta_1 +  \frac{\alpha^3}{\pi^2} \beta_2 +
\frac{\alpha^4}{\pi^3} \beta_3 + \frac{\alpha^5}{\pi^4} \beta_4 + O(\alpha^6), \nonumber\\
&& \widetilde\gamma(\alpha) = \frac{\alpha}{\pi} \gamma_1  + \frac{\alpha^2}{\pi^2} \gamma_2 +
\frac{\alpha^3}{\pi^3} \gamma_3 + O(\alpha^4),
\end{eqnarray}

\noindent
where $\beta_1$, $\beta_2$, $\beta_3$, $\beta_4$, $\gamma_1$, $\gamma_2$, and $\gamma_3$ are real coefficients. A change of a subtraction scheme corresponds to a finite renormalization of the form (\ref{eq:fin_ren}). The functions $\alpha^\prime\left(\alpha\right)$ and $z\left(\alpha\right)$ entering Eq. (\ref{eq:fin_ren}) can be presented as power series in $\alpha/\pi$,

\begin{eqnarray}\label{eqs:loop_fin_ren}
&& \frac{\alpha^\prime(\alpha)}{\pi} = \frac{\alpha}{\pi} + \frac{\alpha^2}{\pi^2} B_2  + \frac{\alpha^3}{\pi^3} B_3 +  \frac{\alpha^4}{\pi^4} B_4 + O(\alpha^5), \nonumber \\
&& z(\alpha) = 1 + \frac{\alpha}{\pi} D_1 + \frac{\alpha^2}{\pi^2} D_2 + O(\alpha^3).
\end{eqnarray}

\noindent
Substituting these expressions into Eqs. (\ref{eq:fin_ren_beta}) and (\ref{eq:fin_ren_gamma}) we obtain the equations describing how the coefficients in Eq. (\ref{eqs:loop_exp_RG}) are transformed under the finite renormalization

\begin{eqnarray}\label{eqs:RG_coeff_fin_ren_beta}
&& \beta_1^\prime = \beta_1; \nonumber\\
&& \beta_2^\prime = \beta_2; \nonumber\\
&& \beta_3^\prime = \beta_3 - B_2 \beta_2 + \left(B_3 - (B_2)^2\right) \beta_1; \nonumber\\
&& \beta_4^\prime = \beta_4 - 2 B_2 \beta_3 + (B_2)^2 \beta_2 + \left(2 B_4 - 6 B_2 B_3 + 4 (B_2)^3\right) \beta_1;\\
\label{eqs:RG_coeff_fin_ren_gamma}
&&\vphantom{1}\nonumber\\
&& \gamma_1^\prime = \gamma_1, \nonumber\\
&& \gamma_2^\prime = \gamma_2 - B_2 \gamma_1 + D_1 \beta_1, \nonumber\\
&& \gamma_3^\prime = \gamma_3 - 2 B_2 \gamma_2 + \left(2 (B_2)^2 - B_3\right) \gamma_1 + D_1 \beta_2 + \left(2 D_2 - (D_1)^2 - 2 B_2 D_1\right) \beta_1.\qquad
\end{eqnarray}

\noindent
From these equations we see that, as is well known, the coefficients $\beta_1$, $\beta_2$, and $\gamma_1$ are scheme-independent. This implies that for the considered theory $\beta_1 = \beta_2 = N_f$ and $\gamma_1 = -1$ remain unchanged under the transformations (\ref{eq:fin_ren_beta}) and (\ref{eq:fin_ren_gamma}). Moreover, all terms in the $\beta$-function proportional to the first power of $N_f$ and all terms in the anomalous dimension without $N_f$ are scheme-independent \cite{Kataev:2013csa} if we assume that the coefficients $B_i$ with $i=2,\ldots$ are proportional to $N_f$.

Let us consider an NSVZ subtraction scheme. The NSVZ relation implies that the coefficients in Eq. (\ref{eqs:loop_exp_RG}) satisfy the equations

\begin{equation} \label{eqs:RG_coeff_NSVZ_cond}
\beta_1 = N_f; \qquad \beta_{i+1} = - N_f \gamma_i.
\end{equation}

\noindent
Then we make the finite renormalization (\ref{eqs:loop_fin_ren}) and require that the NSVZ relation is satisfied after it, so that

\begin{equation} \label{eqs:RG_coeff_NSVZ_cond_new}
\beta_1^\prime = N_f; \qquad \beta_{i+1}^\prime = - N_f \gamma_i^\prime.
\end{equation}

\noindent
Substituting Eqs. (\ref{eqs:RG_coeff_fin_ren_beta}) and (\ref{eqs:RG_coeff_fin_ren_gamma}) into these equalities and taking into account Eq. (\ref{eqs:RG_coeff_NSVZ_cond}), we obtain

\begin{eqnarray}
&& B_3 = (B_2)^2 - N_f D_1, \label{eqs:B_3-B_4}\nonumber\\
&& B_4 = (B_2)^3 - 2 N_f B_2 D_1 - N_f D_2 +\frac{1}{2} N_f (D_1)^2.
\end{eqnarray}

\noindent
Thus, $B_3$ and $B_4$ can be expressed in terms of the other coefficients in Eq. (\ref{eqs:loop_fin_ren}) which remain unfixed and can take arbitrary values.

Let us explicitly verify that Eq. (\ref{eq:NSVZ_cons_fin_ren}) is valid in the considered approximation. For this purpose we substitute the functions $\alpha^\prime\left(\alpha\right)$ and $z\left(\alpha\right)$ given by Eq. (\ref{eqs:loop_fin_ren}) into the left hand side of Eq. (\ref{eq:NSVZ_cons_fin_ren}),

\begin{eqnarray}
&& \frac{1}{\alpha^\prime(\alpha)} - \frac{1}{\alpha} -
\frac{N_f}{\pi}\ln z(\alpha) = \frac{1}{\pi}\Bigg( - B_2 +
\frac{\alpha}{\pi} \Big((B_2)^2 - B_3\Big) + \frac{\alpha^2}{\pi^2} \Big( - (B_2)^3 + 2 B_2 B_3 - B_4\Big) \qquad\nonumber\\
&& - \frac{\alpha N_f}{\pi} D_1  + \frac{\alpha^2 N_f}{\pi^2} \Big(\frac{1}{2}(D_1)^2 - D_2\Big) \Bigg)
+ O(\alpha^3).
\end{eqnarray}

\noindent
Using Eqs. (\ref{eqs:B_3-B_4}) in the right hand side we rewrite this equation as

\begin{equation}
\frac{1}{\alpha^\prime(\alpha)} - \frac{1}{\alpha} - \frac{N_f}{\pi}\ln{z(\alpha)} = - \frac{B_2}{\pi} + O(\alpha^3).
\end{equation}

\noindent
Up to the terms of the second order in $\alpha$ this equation is in agreement with Eq. (\ref{eq:NSVZ_cons_fin_ren}), where

\begin{equation}
B = - \frac{B_2}{\pi}.
\end{equation}

\noindent
Therefore, in the considered approximation the conditions (\ref{eqs:B_3-B_4}) are necessary and sufficient to keep the form of the NSVZ relation. Thus, the arbitrary finite renormalization which does not break the NSVZ equation (\ref{eq:NSVZ_equation}) has the form

\begin{eqnarray}\label{eqs:loop_NSVZ_cons_fin_ren}
&& \frac{\pi}{\alpha^\prime(\alpha)} = \frac{\pi}{\alpha} - B_2 + \frac{\alpha N_f}{\pi} D_1  - \frac{\alpha^2 N_f}{\pi^2} \Big(\frac{1}{2} (D_1)^2 - D_2\Big) + O(\alpha^3), \qquad\nonumber \\
&& z(\alpha) = 1 + \frac{\alpha}{\pi} D_1 + \frac{\alpha^2}{\pi^2} D_2 + O(\alpha^3),
\end{eqnarray}

\noindent
where the coefficients $B_2$, $D_1$, and $D_2$ can take arbitrary values.

\section{Changing the renormalization scale in the three-loop approximation}
\hspace*{\parindent}\label{sec:three_loop_mu_change}

Eqs. (\ref{eq:mu_chng_alpha_sol}) and (\ref{eq:mu_chng_z_sol}) give the finite renormalization corresponding to changing the renormalization point $\mu$. Let us write it explicitly in the lowest orders and verify the general statements discussed above.

We will construct this finite renormalization in the form (\ref{eqs:loop_fin_ren}). Let us start with Eq. (\ref{eq:mu_chng_alpha_sol}), in which we substitute the expansion of $\pi/\widetilde\beta$ in powers of the coupling constant obtained with the help of Eq. (\ref{eqs:loop_exp_RG}),

\begin{equation}
\frac{\pi}{\widetilde\beta(\alpha)} = \frac{1}{\beta_1} \left\{\frac{\pi^2}{\alpha^2} -
\frac{\pi}{\alpha}\, \frac{\beta_2}{\beta_1} + \Big(\frac{{\beta_2}^2}{{\beta_1}^2} -
\frac{\beta_3}{\beta_1}\Big) +  \frac{\alpha}{\pi} \Big(\frac{2\beta_2 \beta_3}{{\beta_1}^2} -
\frac{{\beta_2}^3}{{\beta_1}^3} - \frac{\beta_4}{\beta_1}\Big) \right\} + O(\alpha^2).
\end{equation}

\noindent
Then, integrating from $\alpha$ to $\alpha^\prime$ gives

\begin{equation}
\qquad
\int\limits_{\alpha}^{\alpha^\prime(\alpha)}
\frac{d\alpha}{\widetilde\beta(\alpha)} = \frac{1}{\beta_1} \Bigg\{- \frac{\pi}{\alpha} -
\frac{\beta_2}{\beta_1} \ln{\frac{\alpha}{\pi}} + \frac{\alpha}{\pi} \Big(\frac{{\beta_2}^2}{{\beta_1}^2} -
\frac{\beta_3}{\beta_1}\Big)  + \frac{\alpha^2}{2\pi^2}\Big(\frac{2\beta_2 \beta_3}{{\beta_1}^2} - \frac{{\beta_2}^3}{{\beta_1}^3} -
\frac{\beta_4}{\beta_1}\Big)
\Bigg\} \Bigg|_{\alpha}^{\alpha^\prime(\alpha)} + O(\alpha^3).\qquad
\end{equation}

\noindent
Substituting into this expression the expansion (\ref{eqs:loop_fin_ren}) for $\alpha^\prime(\alpha)$ from Eq. (\ref{eq:mu_chng_alpha_sol}) we obtain

\begin{eqnarray}
\ln{\frac{\mu^\prime}{\mu}} &=& \frac{1}{\beta_1}\Bigg\{B_2 - \frac{\alpha}{\pi} \Big((B_2)^2 - B_3\Big)
 - \frac{\alpha^2}{\pi^2} \Big(2 B_2 B_3 - B_4 - (B_2)^3\Big)  \nonumber \\
&&  - \left(\frac{\alpha}{\pi} B_2  + \frac{\alpha^2}{\pi^2} \Big(B_3 - \frac{1}{2} (B_2)^2\Big) \right)\frac{\beta_2}{\beta_1} +
\frac{\alpha^2}{\pi^2} B_2 \Big(\frac{{\beta_2}^2}{{\beta_1}^2} - \frac{\beta_3}{\beta_1}\Big)\Bigg\}  + O(\alpha^3).
\end{eqnarray}

\noindent
Because this equality should be valid for all values of the coupling constant, it is possible to equate the coefficients of different powers of the coupling constant $\alpha$. This gives the following values of the coefficients in Eq. (\ref{eqs:loop_fin_ren}):

\begin{eqnarray}\label{eqs:B2}
&& B_2 = \beta_1 \ln{\frac{\mu^\prime}{\mu}} \\
\label{eqs:B3}
&& B_3 = (B_2)^2 + \frac{\beta_2}{\beta_1} B_2 \\
\label{eqs:B4}
&& B_4 = (B_2)^3 + \frac{5\beta_2}{2\beta_1} (B_2)^2 + \frac{\beta_3}{\beta_1} B_2.
\end{eqnarray}

Let us proceed to constructing the finite renormalization for the matter superfields. Presenting the integrand in Eq. (\ref{eq:mu_chng_z_sol}) in the form

\begin{equation}
\frac{\widetilde\gamma(\alpha)}{\widetilde\beta(\alpha)} = \frac{1}{\pi} \Bigg\{
\frac{\pi}{\alpha}\, \frac{\gamma_1}{\beta_1}  + \Big(\frac{\gamma_2}{\beta_1} -
\frac{\gamma_1 \beta_2}{{\beta_1}^2}\Big) + \frac{\alpha}{\pi} \Big(\frac{\gamma_3}{\beta_1} -
\frac{\gamma_2 \beta_2}{{\beta_1}^2} - \frac{\gamma_1 \beta_3}{{\beta_1}^2} +
\frac{\gamma_1 {\beta_2}^2}{{\beta_1}^3} \Big) \Bigg\} + O(\alpha^2)
\end{equation}

\noindent
with the help of Eqs. (\ref{eqs:loop_exp_RG}) and taking the integral, we obtain

\begin{eqnarray}
\int\limits_{\alpha}^{\alpha^\prime(\alpha)}
\frac{\widetilde\gamma(\alpha)}{\widetilde\beta(\alpha)} d\alpha &=&
\Bigg\{\frac{\gamma_1}{\beta_1}\ln{\frac{\alpha}{\pi}} + \frac{\alpha}{\pi} \Big(\frac{\gamma_2}{\beta_1} -
\frac{\gamma_1 \beta_2}{{\beta_1}^2}\Big)  + \nonumber \\
&& + \frac{\alpha^2}{2\pi^2} \Big(\frac{\gamma_3}{\beta_1} - \frac{\gamma_2 \beta_2}{{\beta_1}^2} -
\frac{\gamma_1 \beta_3}{{\beta_1}^2} + \frac{\gamma_1 {\beta_2}^2}{{\beta_1}^3}\Big)
\Bigg\} \Bigg|_{\alpha}^{\alpha^\prime(\alpha)} +
O(\alpha^3).
\end{eqnarray}

\noindent
Substituting the expansion of $\alpha^\prime(\alpha)$ given by Eq. (\ref{eqs:loop_fin_ren}) and the value of $B_3$ from Eq. (\ref{eqs:B3}), it is possible to find $\ln z(\alpha)$ from Eq. (\ref{eq:mu_chng_z_sol}),

\begin{equation}
\ln z(\alpha) = \frac{\alpha}{\pi} B_2 \frac{\gamma_1}{\beta_1} + \frac{\alpha^2}{\pi^2} B_2 \left(\frac{\gamma_1}{2\beta_1} B_2 + \frac{\gamma_2}{\beta_1} \right)  +
O(\alpha^3).
\end{equation}

\noindent
Therefore, Eqs. (\ref{eq:mu_chng_alpha_sol}) and (\ref{eq:mu_chng_z_sol}) in the considered order in the coupling constant can be explicitly written as

\begin{eqnarray}\label{eqs:mu_chng_fin_ren}
&& \frac{\pi}{\alpha^\prime(\alpha)} = \frac{\pi}{\alpha} - B_2 -
\frac{\alpha}{\pi} B_2 \frac{\beta_2}{\beta_1}  -  \frac{\alpha^2}{\pi^2} B_2 \left(\frac{\beta_2}{2\beta_1} B_2 +
\frac{\beta_3}{\beta_1} \right) + O(\alpha^3),
\nonumber\\
&& z(\alpha) = 1 + \frac{\alpha}{\pi} B_2 \frac{\gamma_1}{\beta_1} +
\frac{\alpha^2}{\pi^2} B_2 \left(\frac{\gamma_1}{2\beta_1} B_2 + \frac{\gamma_2}{\beta_1} +
\frac{{\gamma_1}^2}{2{\beta_1}^2} B_2\right) +
O(\alpha^3).
\end{eqnarray}

\noindent
Note that so far we do not assume that RGFs satisfy the NSVZ relation.

The transformation (\ref{eqs:mu_chng_fin_ren}) corresponds to changing the renormalization point from $\mu$ to $\mu^\prime$ and is parameterized by $B_2 = \beta_1 \ln{\left(\mu^\prime/\mu\right)}$. We see that the right hand sides of these equations contain the coefficients $\beta_3$ and $\gamma_2$, coming from the the three-loop $\beta$-function and the two-loop anomalous dimension, respectively (see Eq. (\ref{eqs:loop_exp_RG})). This implies that the finite renormalization (\ref{eqs:mu_chng_fin_ren}) is different for different subtraction schemes. Using Eqs. (\ref{eqs:RG_coeff_fin_ren_beta}) and (\ref{eqs:RG_coeff_fin_ren_gamma}) it is also possible to see that under the transformation (\ref{eqs:mu_chng_fin_ren}) $\beta_3' = \beta_3$;\ $\beta_4' = \beta_4$;\ $\gamma_2'=\gamma_2$;\ $\gamma_3' = \gamma_3$, so that in the considered approximation RGFs remain unchanged.

Earlier we argued that the finite renormalization corresponding to the changing of $\mu$ (which is given by Eqs. (\ref{eq:mu_chng_alpha_sol}) and (\ref{eq:mu_chng_z_sol})) does not change the NSVZ relation, but shifts the constant $B$ in Eq. (\ref{eq:NSVZ_cons_fin_ren}). To verify these facts explicitly in the lowest orders, we notice that for the finite renormalization (\ref{eqs:mu_chng_fin_ren})

\begin{eqnarray}
&& \frac{1}{\alpha^\prime(\alpha)} - \frac{1}{\alpha} -
\frac{N_f}{\pi}\ln{z\left(\alpha\right)} = \frac{1}{\pi} B_2 \Big(- 1 -
\frac{\alpha}{\pi}\cdot \frac{\beta_2 + N_f \gamma_1}{\beta_1}  - \nonumber \\
&&\qquad\qquad\qquad\qquad\qquad\qquad - \frac{\alpha^2}{\pi^2} B_2\cdot \frac{\beta_2 + N_f \gamma_1}{2\beta_1}  -
\frac{\alpha^2}{\pi^2}\cdot \frac{\beta_3 + N_f \gamma_2}{\beta_1} \Big) +
O(\alpha^3). \qquad
\end{eqnarray}

\noindent
For NSVZ schemes this equation can be simplified with the help of the NSVZ relation. Namely, using Eq. (\ref{eqs:RG_coeff_NSVZ_cond}) we see that the right hand side does not depend on the coupling constant and is equal to the constant

\begin{equation}
B = - \frac{B_2}{\pi} = - \frac{N_f}{\pi} \ln{\frac{\mu^\prime}{\mu}}
\end{equation}

\noindent
with the considered accuracy (up to terms of the second order in $\alpha/\pi$). This exactly agrees with Eqs. (\ref{eq:NSVZ_cons_fin_ren}) and (\ref{eq:NSVZ_B_hide}). Therefore, the finite renormalization (\ref{eqs:mu_chng_fin_ren}) corresponding to changing of $\mu$ really does not break the NSVZ relation (\ref{eq:NSVZ_equation}) and is parameterized by the constant $B$.

\section{Relation between the NSVZ schemes obtained with DRED and HD in the three-loop approximation}
\hspace*{\parindent}\label{sec:examples}

In the literature NSVZ schemes have been constructed by two main approaches. The first one proposed in \cite{Jack:1996vg,Jack:1996cn} is to make the calculation with dimensional reduction in the $\overline{\mbox{DR}}$-scheme and after it make a specially tuned finite redefinition of the coupling constant (without a finite renormalization of the matter superfields). The coupling constant in the NSVZ scheme constructed in this way we will denote by $\alpha_{\mbox{\scriptsize DR}}$. RGFs in this NSVZ scheme we will denote by the subscript DR and the superscript NSVZ. The second approach is to use the higher derivative regularization and the boundary conditions (\ref{eqs:bc_HD_NSVZ}). The corresponding coupling constant and RGFs will be denoted by the subscript HD. For RGFs we will also write the superscript NSVZ. These two NSVZ schemes are different and can be related by a finite renormalization which satisfy Eq. (\ref{eq:NSVZ_cons_fin_ren}). Let us verify this by explicit calculation in the three-loop approximation.

First, we present the expressions for the three-loop $\beta$-function and the two-loop anomalous dimension of the matter superfields in the NSVZ schemes for ${\cal N}=1$ SQED. If the NSVZ scheme is constructed with dimensional reduction by the help of the above described algorithm, they are written as \cite{Jack:1996vg}

\begin{eqnarray}\label{eqs:RGfunc_DR_NSVZ}
&& \widetilde\beta_{\mbox{\scriptsize DR}}^{\mbox{\scriptsize NSVZ}}(\alpha_{\mbox{\scriptsize DR}}) =
\frac{\alpha_{\mbox{\scriptsize DR}}^2 N_f}{\pi}\Big(1 + \frac{\alpha_{\mbox{\scriptsize DR}}}{\pi} -
\frac{\alpha_{\mbox{\scriptsize DR}}^2}{2\pi^2}(1 + N_f) + O(\alpha_{\mbox{\scriptsize DR}}^3)\Big), \quad\nonumber \\
&& \widetilde\gamma_{\mbox{\scriptsize DR}}^{\mbox{\scriptsize NSVZ}}(\alpha_{\mbox{\scriptsize DR}}) =
- \frac{\alpha_{\mbox{\scriptsize DR}}}{\pi} + \frac{\alpha_{\mbox{\scriptsize DR}}^2}{2\pi^2}(1 + N_f)
+ O(\alpha_{\mbox{\scriptsize DR}}^3)
\end{eqnarray}

\noindent
and, evidently, (by construction) satisfy the NSVZ relation.

In the case of using the higher derivative regularization and the boundary conditions (\ref{eqs:bc_HD_NSVZ}) RGFs take the form \cite{Kataev:2013csa,Kataev:2014gxa}

\begin{eqnarray} \label{eqs:RGfunc_HD_NSVZ}
&& \widetilde\beta_{\mbox{\scriptsize HD}}^{\mbox{\scriptsize NSVZ}}(\alpha_{\mbox{\scriptsize HD}}) =
\frac{\alpha_{\mbox{\scriptsize HD}}^2 N_f}{\pi}\Big(1 + \frac{\alpha_{\mbox{\scriptsize HD}}}{\pi} -
\frac{\alpha_{\mbox{\scriptsize HD}}^2}{2\pi^2}  - \frac{\alpha_{\mbox{\scriptsize HD}}^2 N_f}{\pi^2}\Big[ 1 + \sum\limits_{I}c_I\ln{a_I}\Big]
+ O(\alpha_{\mbox{\scriptsize HD}}^3)\Big), \quad\nonumber\\
&& \widetilde\gamma_{\mbox{\scriptsize HD}}^{\mbox{\scriptsize NSVZ}}(\alpha_{\mbox{\scriptsize HD}}) = - \frac{\alpha_{\mbox{\scriptsize HD}}}{\pi} +
\frac{\alpha_{\mbox{\scriptsize HD}}^2}{2\pi^2} + \frac{\alpha_{\mbox{\scriptsize HD}}^2 N_f}{\pi^2}\Big[ 1 +
\sum\limits_{I}c_I\ln{a_I}\Big] + O(\alpha_{\mbox{\scriptsize HD}}^3)
\end{eqnarray}

\noindent
and also satisfy the NSVZ relation. Here the index $I$ numerates the Pauli--Villars determinants. The coupling constant independent parameters $a_I$ are ratios of the Pauli--Villars masses $M_I$ and the dimensionful parameter $\Lambda$ in the higher derivative term, $a_I = M_I/\Lambda$. The coefficients $c_I$ such that

\begin{equation}\label{eq:c_I_conditions}
\sum_I c_I = 1\qquad \mbox{and} \qquad \sum_Ic_{I}M_I^2 = 0
\end{equation}

\noindent
are related to the powers of the Pauli--Villars determinants in the generating functional and also do not depend on the coupling constant. Eq. (\ref{eq:c_I_conditions}) is the only restriction on the values of $c_I$.

The coefficients $a_I$ can take arbitrary values. In particular, it is possible to choose such $a_I$ and $c_I$ that $\sum_I c_I \ln a_I = -1/2$. In this case RGFs (\ref{eqs:RGfunc_HD_NSVZ}) coincide with (\ref{eqs:RGfunc_DR_NSVZ}). However, below, for completeness, we will not fix values of these parameters.

The finite renormalization relating two above described NSVZ schemes in the lowest order in the coupling constant was constructed in Ref. \cite{Aleshin:2016rrr} and has the form

\begin{eqnarray}\label{eqs:fin_ren_DR_NSVZ-HD_NSVZ}
&& z(\alpha_{\mbox{\scriptsize DR}}) = 1 - \frac{\alpha_{\mbox{\scriptsize DR}}}{\pi} z_1 +
O(\alpha_{\mbox{\scriptsize DR}}^2), \nonumber\\
&& \frac{1}{\alpha_{\mbox{\scriptsize HD}}} = \frac{1}{\alpha_{\mbox{\scriptsize DR}}} - \frac{N_f}{\pi} \Big(\frac{1}{2} +
\sum\limits_{I} c_I\ln{a_I} + z_1\Big) -
\frac{\alpha_{\mbox{\scriptsize DR}} N_f}{\pi^2} z_1 + O(\alpha_{\mbox{\scriptsize DR}}^2),
\end{eqnarray}

\noindent
where $z_1$ is an arbitrary constant. In this case

\begin{equation}
\frac{1}{\alpha_{\mbox{\scriptsize HD}}} - \frac{1}{\alpha_{\mbox{\scriptsize DR}}} - \frac{N_f}{\pi}\ln z(\alpha_{\mbox{\scriptsize DR}}) = - \frac{N_f}{\pi}\Big(\frac{1}{2} +
\sum\limits_{I} c_I\ln{a_I} + z_1\Big) + O(\alpha_{\mbox{\scriptsize DR}}^2).
\end{equation}

\noindent
In agreement with Eq. (\ref{eq:NSVZ_cons_fin_ren}) the right hand side of this equation is a constant ($B$) up to the terms of the second order in the coupling constant.

The arbitrary constant $z_1$ originates from the arbitrariness in choosing the renormalization point. Really, the constant $z_1$ can be removed from the equations (\ref{eqs:fin_ren_DR_NSVZ-HD_NSVZ}) by the finite renormalization

\begin{eqnarray}\label{eqs:fin_ren_no-z1}
&& z^\prime(\alpha_{\mbox{\scriptsize HD}}) = 1 + \frac{\alpha_{\mbox{\scriptsize HD}}}{\pi}  z_1 +
O(\alpha_{\mbox{\scriptsize HD}}^2), \nonumber\\
&& \frac{\pi}{\alpha^\prime} = \frac{\pi}{\alpha_{\mbox{\scriptsize HD}}} + N_f z_1 +
\frac{\alpha_{\mbox{\scriptsize HD}} N_f}{\pi} z_1 + O(\alpha_{\mbox{\scriptsize HD}}^2),
\end{eqnarray}

\noindent
which has the form (\ref{eqs:mu_chng_fin_ren}) with $B_2 = -z_1 N_f$. In other words, this transformation corresponds to the change of the renormalization point $\mu \to \mu^\prime = \mu e^{- z_1}$.

Thus, the finite renormalization connecting two above described NSVZ renormalization schemes really satisfies Eq. (\ref{eq:NSVZ_cons_fin_ren}) and contains an arbitrary constant corresponding to the arbitrariness in choosing the renormalization scale $\mu$.

Finally, we note that the finite constant $x_0$ (which is a certain fixed value of $\ln\Lambda/\mu$) is also present in the boundary conditions (\ref{eqs:bc_HD_NSVZ}). Evidently, changing this constant is equivalent to changing the renormalization scale $\mu$,

\begin{equation}
\mu\to \mu' = \exp(x_0-x_0')\,\mu.
\end{equation}

\noindent
Therefore, the corresponding finite renormalization is given by Eqs. (\ref{eq:mu_chng_alpha_sol}) and (\ref{eq:mu_chng_z_sol}) and satisfy Eq. (\ref{eq:NSVZ_cons_fin_ren}) with

\begin{equation}
B = \frac{N_f}{\pi}(x_0' - x_0).
\end{equation}

\section{Conclusion}
\hspace*{\parindent}

The NSVZ relation (\ref{eq:NSVZ_equation}) for RGFs defined in terms of the renormalized coupling constant is valid only in special subtraction schemes. In particular, if the regularization is made by the dimensional reduction method and the renormalization is made by the $\overline{\mbox{DR}}$ prescription, then the NSVZ equation is obtained only after a specially tuned redefinition of the coupling constant. In the case of using the higher covariant derivative regularization for ${\cal N} = 1$ SQED with $N_f$ flavors the NSVZ renormalization scheme is constructed in all loops by imposing the boundary conditions (\ref{eqs:bc_HD_NSVZ}). However, this NSVZ scheme is different from the NSVZ scheme constructed with dimensional reduction by the above described method, because RGFs are different (see Eqs. (\ref{eqs:RGfunc_DR_NSVZ}) and (\ref{eqs:RGfunc_HD_NSVZ})). This implies that the NSVZ scheme is not unique.

In this paper we demonstrate that the NSVZ relation (\ref{eq:NSVZ_equation}) is valid for a class of subtraction schemes connected by finite renormalizations of a special form. Namely, the functions $\alpha^\prime(\alpha)$ and $z(\alpha)$ giving the finite renormalizations of the coupling constant and of the matter superfields, respectively, should be related by Eq. (\ref{eq:NSVZ_cons_fin_ren}). The constant $B$ entering this equation appears due to arbitrariness in choosing the renormalization point $\mu$. Changing a value of $\mu$ corresponds to changing the constant $B$. This implies that the constant $B$ in Eq.
(\ref{eq:NSVZ_cons_fin_ren}) can be removed by a proper redefinition of $\mu$, see Eq. (\ref{eq:new-mu}). Thus, if the subtraction schemes which differ only by a value of $\mu$ are considered as equivalent, then the class of the NSVZ schemes is parameterized by a single finite function (e.g., $z(\alpha)$).

The results listed above are explicitly verified in the three-loop approximation. In particular, an arbitrary finite renormalization which does not break the NSVZ relation (\ref{eq:NSVZ_equation}) can be written in the form (\ref{eqs:loop_NSVZ_cons_fin_ren}).

\section*{Acknowledgements}
\hspace*{\parindent}

The work of I.G. and A.K. was supported by the Foundation for the Advancement
of Theoretical Physics and Mathematics ``BASIS'', grant No. 17-11-120.

\end{document}